\documentclass[iop]{emulateapj}
\usepackage{epsfig}
\usepackage{url}

\newcommand{\grb}{GRB~110625A}
\newcommand{\gr}{$\gamma$-ray}
\newcommand{\grs}{$\gamma$-rays}

\shorttitle{Fermi LAT observations of GRB~110625A}
\shortauthors{Tam et al.}

\begin{document}

\title{Fermi Large Area Telescope observations of GRB~110625A}

\author{P.H.T.~Tam\altaffilmark{1}, A.K.H.~Kong\altaffilmark{1,2}, and Yi-Zhong Fan\altaffilmark{3,4}
}

\affil{$^1$ Institute of Astronomy and Department of Physics, National Tsing Hua University, Hsinchu 30013, Taiwan\\
$^2$ Golden Jade Fellow of Kenda Foundation, Taiwan\\
$^3$ Purple Mountain Observatory, Chinese Academy of Sciences, Nanjing 210008, China\\
$^4$ Key Laboratory of Dark Matter and Space Astronomy, Chinese Academy of Sciences, Nanjing
210008, China}
\email{phtam@phys.nthu.edu.tw}

\begin{abstract}
Gamma-ray bursts (GRBs) that emit photons at GeV energies form a small but significant population of GRBs. However, the number of GRBs whose GeV-emitting period is simultaneously observed in X-rays remains small. We report \gr~observations of \grb~using Fermi's Large Area Telescope (LAT) in the energy range 100~MeV--20~GeV. Gamma-ray emission at these energies was clearly detected using data taken between 180~s and 580~s after the burst, an epoch after the prompt emission phase. The GeV light curve differs from a simple power-law decay, and probably consists of two emission periods.
Simultaneous Swift/XRT observations did not show flaring behaviors as in the case of GRB~100728A. We discuss the possibility that the GeV emission is the synchrotron self-Compton radiation of underlying ultraviolet flares.
\end{abstract}


\section{Introduction}

Since the launch of the Fermi satellite in 2008, more than 20 \gr~bursts (GRBs) have been detected above $\sim$100~MeV by the Large Area Telescope (LAT) aboard the satellite~\citep{lat_090510,lat_080825c,lat_080916c,lat_090902b}. Long-lived MeV--GeV emission of GRBs, first detected in the EGRET era, is now a common feature of LAT-detected GRBs. The nature of such temporally extended emission beyond the prompt GRB phase is not well understood. A widely-discussed radiation mechanism is synchrotron emission from external shocked electrons~\citep[e.g.,][]{Zou09,BarniolDuran09,Ghisellini10}, but inverse-Compton scattering off flare photons or late-time activities of the central engine are among alternative scenarios~\citep{lat_100728a,Zhang11}.

Simultaneous observations at other wavelengths of such extended MeV--GeV emission from GRBs are crucial to disentangle various emission models. By May 2011, only two LAT-detected GRBs have been simultaneously observed by \emph{Swift}'s X-ray telescope (XRT) during its GeV-emitting period: GRB~090510 and GRB~100728A. GRB~090510 remains the only short GRB detected by LAT, and its GeV emission can be interpreted as, e.g., synchrotron radiation of the forward shock electrons~\citep[][but see \citet{Gao2009} and \citet{Piran_Nakar10} for an opposing viewpoint]{090510_afterglow,Ghirlanda_090510}. In the case of GRB~100728A, X-ray flares and corresponding GeV emission were detected up to $\sim$1~ks after the burst, suggesting their common origin~\citep{lat_100728a}. However, the number of GRBs whose GeV emission is simultaneously covered in X-rays remains low.

In this paper, we report another such case: \grb, that was detected by Fermi/LAT and Swift/XRT simultaneously for several hundred seconds. Errors are reported at 1$\sigma$, unless otherwise specified.

\section{GRB~110625A}

At 21:08:28 UT on 2011 June 25, the Burst Alert Telescope (BAT) aboard \emph{Swift} triggered on GRB~110625A~\citep{page12088}. The refined BAT position was R.A.~$=19\mathrm{^h}07\mathrm{^m}00\fs3$, Dec.~$=+06\arcdeg45\arcmin17\farcs8$ (J2000) with an uncertainty of 1\farcm3~\citep[90\% containment radius;][]{palmer12091}. The BAT light curve showed a multiple-peaked structure lasting from $\sim-12\,\mathrm{s}$ to $\sim18\,\mathrm{s}$ with a tail extending up to $\sim$150~s with respect to BAT trigger time. $T_{90}$ (referred to the time interval between the instants at which 5\% and 95\% of the total integral emission is detected in the 15--350 keV band) was $44.5\pm10.1\,\mathrm{s}$~\citep{palmer12091}. Konus-Wind~\citep{golenetskii12093}, Fermi/GBM, INTEGRAL~SPI-ACS~\citep{gruber12100}, and Suzaku/WAM~\citep{mizuno12102} also triggered on GRB~110625A. The Konus-Wind team reported that the 20~keV to 10~MeV time-averaged spectrum from 0 to $58.88\,\mathrm{s}$ after the Konus-Wind trigger time is best fitted by the Band function with $\alpha=-1.05\pm0.08$, $\beta=-2.7^{+0.2}_{-0.5}$, and $E_\mathrm{p}=190^{+17}_{-14}$~keV; emission was seen up to $\sim$8~keV. The burst fluence is $(6.1\pm0.6)\times10^{-5}$~erg~cm$^{-2}$~\citep{golenetskii12093}.

Fermi/GBM triggered on GRB~110625A at 21:08:18.24 UT ($T_\mathrm{0}$) on 2011 June 25. The angle of the GRB position is 88$^\circ$ from the LAT boresight at $T_\mathrm{0}$. The autonomous rapid repoint maneuver repointed the LAT such that GRB~110625A was put in the field-of-view (FoV) of LAT from $\sim T_\mathrm{0}+100$ to $T_\mathrm{0}+600$. However, due to the poorly measured position derived on board by GBM (off by $\sim$68$\arcdeg$), The burst position was placed at the outskirt of the FoV of Fermi/LAT, diminishing its sensitivity~\citep{gruber12100}.


\emph{Swift}/XRT began data-taking of the burst at $\approx T_\mathrm{0}+150$~s and found the X-ray afterglow source at the position R.A.~$=19\mathrm{^h}06\mathrm{^m}55\fs85$, Dec.~$=+6\arcdeg42\arcmin19\farcs2$~\citep[J2000;][]{page12092} with an error circle of radius~$2\farcs1$ (90\% confidence level). This position is used in analyses presented in this paper. The flux faded initially as a power law with an index of $\alpha_1=1.14 \pm 0.04$ and then as $\alpha_2=2.3^{+1.6}_{-0.4}$, after a break at $17^{+11}_{-10}$~ks after the BAT trigger~\citep[90\% confidence level;][]{page_gcnr_336}.

The XRT observations of \grb~started in Window Timing (WT) mode that lasted for $\approx$90~s. Then data were taken in the Photon counting (PC) mode since $\sim T_\mathrm{0}+250$~s. The WT(PC) spectrum can be fit by an absorbed power-law model with photon index $2.5\pm0.4$($1.8\pm0.4$)~\citep[90\% confidence level;][]{page12092}, suggesting a marginal hardening between early and late times. We tried the absorbed blackbody model as well, finding no improvement on the fit on both WT- and PC-mode data. Using PC-mode data taken from $T_\mathrm{0}+255$~s to $T_\mathrm{0}+580$~s (to match LAT observations), we fit the time-averaged 0.3--10~keV energy spectrum with an absorbed power-law model. A photon index of $1.44^{+0.23}_{-0.22}$ and a hydrogen column density of $4.53^{+0.63}_{-0.59}\times 10^{22} \mathrm{cm}^{-2}$ were obtained. In our analysis, we used C-statistic during the fitting process. Piled-up effects are present in the PC-mode data concerned and were removed by ignoring the circular region with radius $7\farcm2$. 

Interestingly  there is no clear early steep decay phase, a common feature in Swift XRT afterglows that has been widely interpreted as the high latitude emission from the prompt (Fenimore et al. 1996; Kumar \& Panaitescu 2000). The absence of a sharp decline phase may be just due to the fact that XRT observations started only at $\approx T_0+150$ s, i.e., the afterglow started earlier. Such a possibility provides us an additional argument to classify the LAT emission as post prompt emission.

In the optical and infrared bands, the only object suspected of being variable in the XRT FoV was reported in~\citet{gorosabel12098}. However, from an archival Canada-France-Hawaii Telescope r-band image (taken on 2010 May 8), the object was detected with $r=23.9\pm0.1$ and not detected with the g-band filter with a 5-sigma detection limit of 26.2, consistent with the GROND observations~\citep{filgas12096}. We therefore conclude that it is unrelated to~\grb.

\section{The Fermi/LAT Data Analysis and Results}

We analyzed the LAT data that are available at the Fermi Science Support Center\footnote{\url{http://fermi.gsfc.nasa.gov/ssc/}}. The Fermi Science Tools v9r23p1 package was used to reduce and analyze the data. We selected photons of energies between 100~MeV and 20~GeV. To reduce the contamination from Earth albedo $\gamma$-rays, we excluded events with zenith angles greater than 100$^\circ$.

\begin{figure*}
    \epsscale{1.}
    \plotone{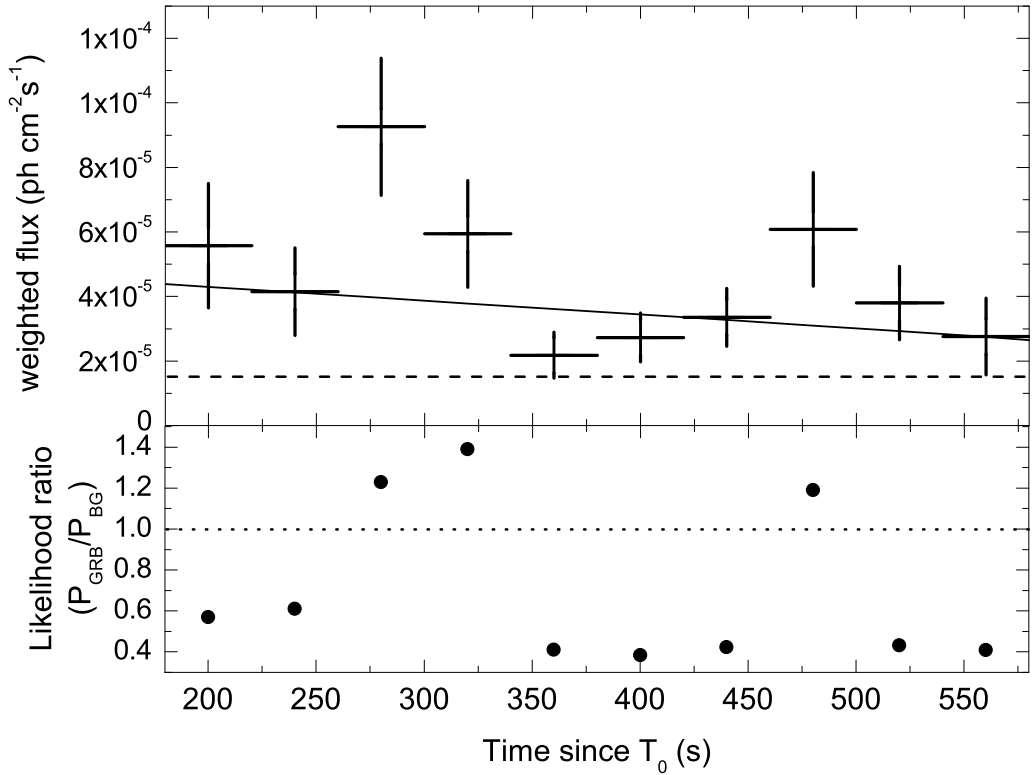}
      \caption{\emph{Top panel}: Weighted photon flux between 100~MeV and 20~GeV from \grb~as observed using Fermi/LAT within a circular region of radius 10$^\circ$, centered at the XRT position, between $T_\mathrm{0}+180$~s and $T_\mathrm{0}+580$~s. Each bin represents 40~s. The solid line represents the best-fit power law, with $\chi^2/$d.o.f.$=16.79/8$. The dashed line indicates the estimated number of background events averaged in the whole period. \emph{Bottom panel}: Ratio, $R:=P_{\rm GRB}/P_{\rm BG}$, of the contribution from \grb~over that from background based on a likelihood test for each time bin. The dotted line stands for $R=1$.}
         \label{lightcurve}
   \end{figure*}

We selected photons from a region-of-interest (ROI) of a 10$^\circ$-radius circular region centered on the XRT position of \grb~and plot them in Fig.~\ref{lightcurve}. Here we use ``P7TRANSIENT'' data so as to increase photon statistics. We selected the time span during which the inclination angle of the GRB position is less than 66$^\circ$ to make sure that the GRB position is well within the LAT FoV, corresponding to $T_\mathrm{0}+180$~s to $T_\mathrm{0}+580$~s. The count rate is normalized by the varying exposure, that was computed by using the best-fit spectral index $\Gamma_\gamma=2.7$ obtained in \emph{gtlike} (described below). The background level of a time bin $i$ was calculated by $P_{\rm BG,i}=\sum_{j=1}^{N_i}b_j$, where
$b_j$ represents the probability that a particular photon $j$ in time bin $i$ comes from the Galactic diffuse emission, i.e., from the background, and $N_i$ the number of photons in the ROI. A similar approach was used by~\citet{BW_pulsation}. This probability is calculated using \emph{gtsrcprob} based on the best likelihood fit between $T_\mathrm{0}+180$~s and $T_\mathrm{0}+580$~s, according to the method of~\citet{kerr11}.

The weighted photon flux from \grb~was calculated by $P_{\rm GRB,i}=\sum_{j=1}^{N_i}w_j$, where $w_j=1-b_j$ represents the probability that the photon originates from the GRB, normalized by the exposure. Thus, the total number of counts, $N_i$, is scaled by the factor $P_{\rm GRB,i}/N_i$ to give rise to the weighted photon counts. The error bars were scaled by the same ratio. The background level for the weighted photon flux is estimated by $B_i=\sum_{j=1}^{N_i}w_j\times b_j$ for each bin and was averaged over the whole period.
It appears that the weighted light curve differs from a simple power law decay. To quantify this, we fit the weighted light curve with a simple power law between $T_\mathrm{0}+180$~s and $T_\mathrm{0}+580$~s: $f=f_0+at$, $t$ measured in seconds, and found the best fit parameters to be $f_0=5.2\times$10$^{-5}$~cm$^{-2}$s$^{-1}$ and $a=-4.3\times$10$^{-8}$~cm$^{-2}$s$^{-2}$, giving $\chi^2/$d.o.f.$=16.79/8$. Thus the light curve deviates from a simple power law decay at the level of 96.77\%.

The ratios $P_{\rm GRB,i}/P_{\rm BG,i}$ are shown in Fig.~\ref{lightcurve} as well. One can see that during $T_\mathrm{0}+260$~s to $T_\mathrm{0}+340$~s and $T_\mathrm{0}+460$~s to $T_\mathrm{0}+500$~s, the contribution from GRB~110625A is higher than the background, since $P_{\rm GRB,i}/P_{\rm BG,i}>1$. As an alternative algorithm, we have also assigned each photon to either \grb~or background depending on whether $w_j>0.5$ (\grb) or $w_j<0.5$ (background) and calculated the ratio of the number of photons from \grb~and those from background. This method gives consistent results.

The existence of two emission epochs from \grb~was first noted in~\citet{tam12097} and characterized by~\citet{gruber12100}. Fitting the light curve with two Gaussian profiles on top of a baseline emission, the peaks and widths of the first and second emission epochs are given by $t_\mathrm{p_1}=T_\mathrm{0}+285.8\pm10.0$~s and $\Delta_1=64.9\pm22.3$~s, and by $t_\mathrm{p_2}=T_\mathrm{0}+483.8\pm15.7$~s and $\Delta_2=59.0\pm31.0$~s, respectively. Therefore, the two emission epochs are roughly $T_\mathrm{0}+220.9$~s to $T_\mathrm{0}+350.7$~s and  $T_\mathrm{0}+424.8$~s to $T_\mathrm{0}+542.8$~s.

We then performed unbinned maximum-likelihood analyses (\emph{gtlike}) of a 15$\degr$-ROI centered at the XRT position to characterize the detection significance and spectrum of the $>$100~MeV \grs~from \grb. Events that are classified as ``P7SOURCE'' and arrived between $T_\mathrm{0}+180$~s and $T_\mathrm{0}+580$~s are used. We subtracted the background contribution by including the Galactic diffuse model (gal\_2yearp7v6\_v0.fits). Contributions from other sources (including the isotropic background) are negligible and not included. The spectral index of the GRB and normalization values of all three components were set free in the likelihood analysis. The likelihood fit returned a best-fit spectral index of $\Gamma_\gamma=2.7\pm0.3$ and
a \emph{test-statistic} (TS) value~\citep{Mattox_96} of 52.9, corresponding to a detection significance of $\sim$7$\sigma$. Using \emph{gtfindsrc}, we found the best-fit LAT position to be right ascension (J2000) $=$ 286$\fdg$51 and declination (J2000) $=$ 6$\fdg$86 with statistical uncertainty of 0$\fdg$44, which is consistent with the XRT position. The systematic uncertainty should be $\ga$0$\fdg$6 based on GRB~080825C which also occurred at large inclination angle at the GRB onset~\citep{lat_080825c}.

\begin{figure*}
    \epsscale{1.}
   \plotone{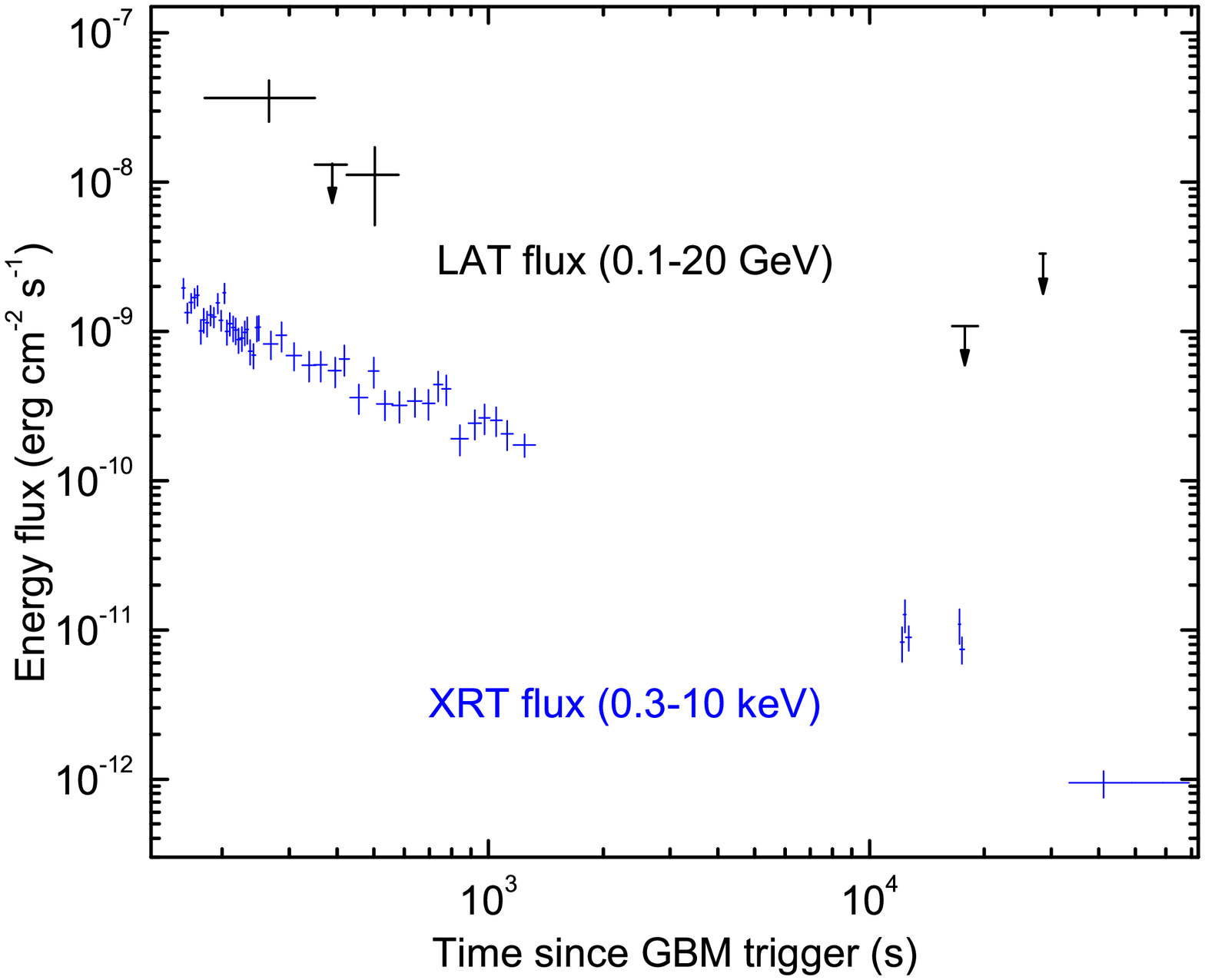}
      \caption{The LAT energy flux derived from unbinned likelihood analyses of \grb~is shown in black. The XRT energy flux is shown in blue for comparison~\citep{xrt_evan07,xrt_evan09}. The X-ray light curve does not show any prominent X-ray flaring activity.}
         \label{flux}
   \end{figure*}

Performing unbinned maximum-likelihood analyses (\emph{gtlike}) of the same 15$\degr$-ROI for different time bins, we produced the background-subtracted light curve, as shown in Figure~\ref{flux}. The first and third data point represents the periods (I) $T_\mathrm{0}+180$~s to $T_\mathrm{0}+350$~s and (III) $T_\mathrm{0}+420$~s to $T_\mathrm{0}+580$~s, respectively. These two periods, (I) and (II), were chosen to cover the two emission epochs characterized above, as well as the short periods $T_\mathrm{0}+180$~s to $T_\mathrm{0}+220.9$~s and $T_\mathrm{0}+542.8$~s to $T_\mathrm{0}+580$~s, respectively. Limited photon statistics do not allow any meaningful analysis on these $\sim$40~s periods. The period in-between (II) during which no emission was detected is plotted as a 90\% confidence-level upper limit. The burst position entered the LAT FoV again after $T_\mathrm{0}+16.5$~ks. No emission was detected and we derived upper limits for these late epochs as well. The \emph{Swift}/XRT light curve is also plotted in the same figure. No flaring behavior is present in the X-ray light curve.

The spectral properties of the GeV emission for different times are summarized in Table~\ref{lat_spec}. There is no significant spectral change between periods (I) and (III), as opposed to preliminary results presented by~\citet{gruber12100}, who found that the spectrum for the second emission period was much softer than the first period (we were able to reproduce the same results by using Pass6 data). Since Pass7 data are recommended after their release, we will take the more conservative proposition that there is no subtle change in the spectral index in the LAT data.

\begin{table}
\centering
\caption{Spectral properties of the GeV emission during different periods. \label{lat_spec}}
\begin{tabular}{ccr}
    \tableline\tableline
    $t-T_\mathrm{0}$ (s) & Photon Index & TS \\ 
    \tableline
    I: 180--350 & $-$2.6$\pm$0.3 & 42.8 \\ 
    II: 350--425 & $...$ & 0 \\ 
    III: 425--580 & $-$2.7$\pm$0.6 & 8.5 \\ 
    180--580 & $-$2.7$\pm$0.3 & 52.9 \\ 
    \tableline
\end{tabular}
\end{table}

\begin{figure*}
    \epsscale{1.}
   \plotone{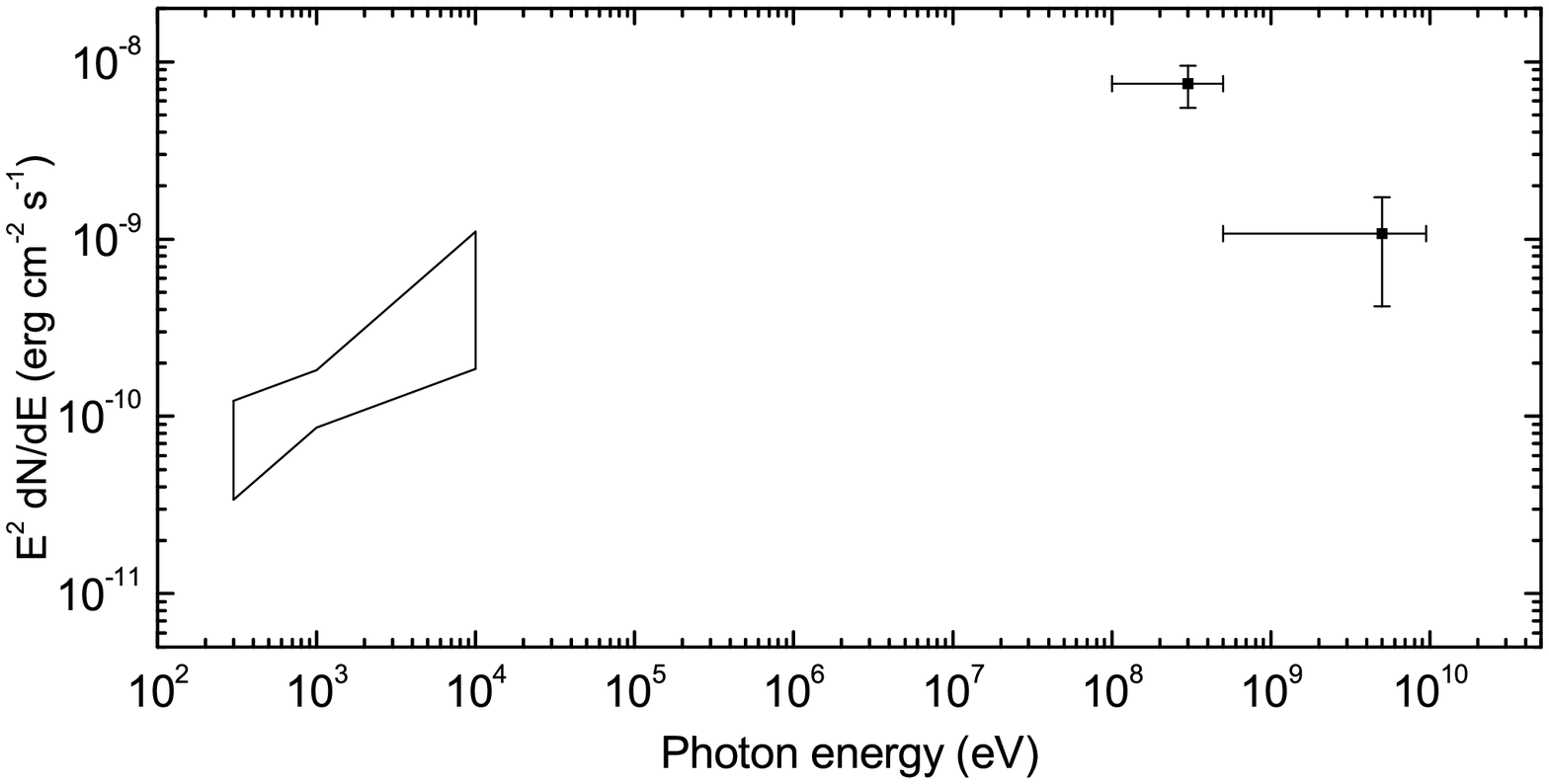}
      \caption{Spectral energy distribution of the X-ray and \gr~emission integrated over the time intervals $T_\mathrm{0}+255$~s to $T_\mathrm{0}+580$~s and $T_\mathrm{0}+220$~s to $T_\mathrm{0}+580$~s, respectively. The butterfly represents the model absorbed power-law spectrum in the 0.3--10~keV range.}
         \label{sed}
   \end{figure*}

Figure~\ref{sed} shows the spectral energy distribution of the X-ray and \gr~emission integrated over the time intervals $T_\mathrm{0}+255$~s to $T_\mathrm{0}+580$~s and $T_\mathrm{0}+220$~s to $T_\mathrm{0}+580$~s, respectively.

%

\section{Discussion}

The redshift of GRB 110625A is unknown and we assume a redshift $z\sim 1$. As for the afterglow, the X-ray spectral index $\sim -0.44\pm 0.22$ and the early decline $t^{-1.14}$ (up to a time $\sim 2\times 10^{4}$ s) can be reproduced if the forward shock is in the slow cooling phase, i.e., $\nu_{\rm m}<\nu_{\rm X}<\nu_{\rm c}$ (where $\nu_{\rm m}$ is the typical synchrotron radiation frequency and $\nu_{\rm c}$ is the cooling frequency), the power-law index of the shock-accelerated electrons $p\sim 2.5$ and the circum-burst medium has a constant number density. In such a scenario,
the X-ray data suggest that $\nu_{\rm c} \geq 10^{18}(t/2\times 10^{4}~{\rm s})^{-1/2}$. In the synchrotron radiation model, the cooling frequency is related to the cooling Lorentz factor $\gamma_{\rm c}$ as $\nu_{\rm c}\approx {m_{\rm e}c \gamma_{\rm c}^{2} \Gamma B /2\pi(1+z)}$, where $m_{\rm e}$ is the rest mass of electrons, $c$ is the speed of light, $\Gamma\approx 88 E_{\rm k,54}^{1/8}
n^{-1/8}(t/300~{\rm s})^{-3/8}[(1+z)/2]^{3/8}$ is the bulk Lorentz factor of the decelerating outflow and $B \approx 3.5~{\rm Gauss}~\epsilon_{\rm B,-2}^{1/2}E_{\rm k,54}^{1/8}n^{3/8}(t/300~{\rm s})^{-3/8}[(1+z)/2]^{3/8}$ is the strength of magnetic field in the emitting region (e.g., Piran 1999). Please note that $E_{\rm k}$ is the kinetic energy of the outflow, $n$ is the number density of the circum-burst medium and $\epsilon_{\rm B}$ is the fraction of the shock energy given to the magnetic field. The convention $Q_{\rm n}=Q/10^{\rm n}$ has been
adopted here and throughout this work. At $t\sim 300$ s, the time of the strong GeV emission, we have $\nu_{\rm c}\geq 8\times 10^{18}$ Hz and the corresponding cooling Lorentz factor $\gamma_{\rm c} \sim 2\times 10^{6}~{(\Gamma B)}^{-1/2}\sim 10^{5}\epsilon_{\rm B,-2}^{-1/4}E_{\rm k,54}^{-1/8}n^{-1/8}$.  To account for the GeV spectrum one needs a typical energy of seed photons as low as $\sim 100~{\rm MeV}/\gamma_{\rm c}^{2}\sim 2.4\times 10^{12}~{\rm Hz}$, which rules out the possibility that the GeV afterglow is dominated by the synchrotron self-Compton (SSC) radiation of the external forward shock. The synchrotron radiation of the forward shock electrons may produce prominent GeV emission \citep[e.g.,][]{Zou09,BarniolDuran09}. As shown in Fig.3, the GeV emission could be the high-energy tail of the synchrotron radiation component as long as the spectrum $F_\nu \propto \nu^{-(p-1)/2}\sim \nu^{-0.75}$ holds in the energy range from 0.2 keV to $\sim 100$ MeV. Such a peculiar spectrum suggests a $\nu_{\rm c} \sim 100$ MeV, requiring $E_{\rm k,54}^{-1/2}\epsilon_{\rm B,-2}^{-3/2}n^{-1}\sim 3\times 10^{6}$, where we have taken the expression of $\nu_{\rm c}$ by Yost et al. (2003). Alternatively, one can imagine that there was a far-infrared flare and the forward shock electrons were cooled by the flare photons, producing a GeV radiation component via the inverse Compton scattering. These two kinds of processes may be able to account for some properties of the GeV emission (e.g., the spectrum and the flux) but possibly not the temporal behavior. As shown in Fig.1, the possible abrupt decline of the GeV emission may be hard to be interpreted within the forward shock scenario.

In the following investigation, we assume that the forward shock synchrotron GeV radiation is unimportant (requiring that $\nu_{\rm c}\ll 100$ MeV, i.e., $E_{\rm k,54}^{-1/2}\epsilon_{\rm B,-2}^{-3/2}n^{-1} \ll 3\times 10^{6}$) and then examine the possibility that the GeV emission is the SSC radiation of an underlying ultraviolet (UV) flare with a peak energy $E_{\rm p,flare}\sim 20$ eV and a spectrum $\propto \nu^{-1.7}$ for $h\nu>E_{\rm p,flare}$ (as already mentioned in Sect.~3, there was no flaring behavior in X-ray band). In the GRB afterglow, optical/ultraviolet flares could be powered via the so-called late internal shocks, as observed for example in GRB 080129 \citep{Greiner09,Gao2009a}.  Adopting eq.(49) of Fan \& Piran (2008), we have $E_{\rm p}^{\rm ssc} \sim  100 {\rm MeV}~\varepsilon_{-4}^{-1/4}L_{\rm flare,49}^{-1/2} R_{\rm
flare,15.5}(E_{\rm p,UV}/20~{\rm eV})^{2}$, matching the data, where $\varepsilon\equiv \epsilon_{\rm B,flare}/\epsilon_{\rm e,flare}$ (where $\epsilon_{\rm e}$ is the fraction of  shock energy given to electrons in the flare phase), and the luminosity of the UV flare is related to the luminosity of GeV emission as $L_{\rm flare} \sim \varepsilon^{1/2} L_{\rm GeV} \sim 3\times 10^{48}\varepsilon_{-4}^{1/2}$ erg/s (where $L_{\rm GeV}\sim 3\times 10^{50}~{\rm erg/s}$ is the luminosity of the GeV emission). The requirement $\varepsilon\ll 1$ suggests that the outflow is baryonic and the magnetic field in the emitting region is shock-generated. The required $R_{\rm flare}\sim 3\times 10^{15}$ cm imposes a tight constraint on the bulk Lorentz factor of the flare outflow, i.e., $\Gamma_{\rm flare}>[(1+z)R_{\rm flare}/160~{\rm s}/3\times 10^{10}~{\rm cm~s^{-1}}]^{1/2}\sim 25$ (The current data do not allow us to perform a reliable estimate on the variability timescale of the GeV emission; that is why we used the whole period (see Fig.1) to constrain $\Gamma_{\rm flare}$). To be a valid interpretation, three more requirements should be satisfied: (i) the flare emission in X-ray band should be lower than the forward shock X-ray emission; (ii) $\gamma_{\rm c}\geq 10^{5}$ still holds in the flare phase, in which the forward shock electrons suffer additional cooling by the UV photons; (iii) the GeV photons are not absorbed via pair production on the high energy tail of the flare or on the SSC MeV photons. The first requirement is satisfied since the observed X-ray luminosity at $t\sim 300$ s is $\sim 3\times 10^{48}$ erg/s while the emission of the UV flare in the X-ray band is $\sim L_{\rm flare}(0.3~{\rm keV}/E_{\rm p,flare})^{-0.7} \sim 10^{48}$ erg/s. The second requirement is met as long as the interstellar medium surrounding the burst has a number density $n<0.01~{\rm cm}^{-3}$ and $\epsilon_{\rm B}<0.01$ (It is straightforward to draw this conclusion with eq.(2) and eq.(5) of Fan \& Piran 2006). Substituting the current physical parameters into eq.(43) of Fan et al. (2008), it is straightforward to show that the absorption by pair production on the high energy tail of the flare can be ignored. The absorption by pair production on the SSC MeV photons can be calculated in a rather similar way and is found to be not serious \footnote{With the SSC spectrum $F_{\rm E}\propto E^{-1/2}$ for $E\leq E_{\rm p}^{\rm ssc}$ and $F_{\rm E}\propto E^{-(\Gamma_\gamma-1)}$ for $E>E_{\rm p}^{\rm ssc}$, we just need to replace the expression of the term $N_{\rm >E_{\rm a}}$ in eq.(43) of Fan et al. (2008) by the new form $8\pi (\Gamma_\gamma-2) D_{\rm L}^2 (E_{\rm a}E_{\rm p}^{\rm ssc})^{-1/2}F_{\rm SSC} \delta t /[(1+z)(2\Gamma_\gamma-3)]$ to estimate the pair production optical depth $\tau_\gamma$, where $F_{\rm SSC}$ is the flux of the SSC radiation component and $\delta t$ is the observed typical variability timescale, $E_{\rm a}\approx 0.5~{\rm MeV}~\Gamma_{1.5}^{2}(E_{\gamma}/1~{\rm GeV})^{-1}$ is the energy of the photons that can absorb the GeV emission by pair production. For $E_{\rm p}^{\rm SSC}\sim 100$ MeV, $E_{\rm a}\sim 0.5$ MeV, $R_{\rm flare}\geq 3\times 10^{15}$ cm, $F_{\rm SSC} \sim 10^{-7}~{\rm erg~s^{-1}~cm^{-2}}$, $\delta t \sim 80$ s, $\Gamma_\gamma \sim 2.7$ and $z\sim 1$, we have $\tau_{\gamma\gamma}\approx 11\sigma_{\rm T}N_{\rm >E_{\rm a}}/720\pi R_{\rm flare}^{2}\sim 0.4$ (e.g. Svensson 1987), where $\sigma_{\rm T}$ is the Thompson cross section.}. In other words, the third requirement is also satisfied. Therefore we conclude that the SSC radiation of an underlying bright UV flare could account for the GeV radiation from $t\sim 260$ s to $340$~s. The weak GeV emission in the time interval $460-500$ s may be interpreted in the same scenario.


\section{Conclusions}

\grb~is the third GRB detected by Fermi/LAT and Swift/XRT simultaneously. We have shown that the GeV light curve differs from a simple power-law decay, and probably consists of two emission periods. The rapid decrease of GeV flux during both periods challenges the notion that the emission comes from the external forward shock. While in the case of GRB~100728A, late-time X-ray flares seem to accompany the GeV emission, no such flares are seen in the time frame during which GeV emission was detected. This suggests a different origin of the GeV emission between the two cases.  We discuss the possibility that the GeV emission is the SSC radiation of an underlying ultraviolet (UV) flare. Multiwavelength coverage of the rare class of LAT GRBs during the GeV-emitting period is crucial.

\acknowledgments
We thank Tsvi Piran and Xiangyu Wang for useful discussion. This research made use of data supplied by the High Energy Astrophysics Science Archive Research Center (HEASARC) at NASA's Goddard Space Flight Center, and the UK Swift Science Data Centre at the University of Leicester. This project is supported by the National Science Council of the Republic of China (Taiwan) through grants NSC100-2628-M-007-002-MY3
and NSC100-2923-M-007-001-MY3. YZF is supported in part by National Basic Research Program of China under grant 2009CB824800 and National Natural Science Foundation of China under grant 11073057, and by the 100 Talents Program of Chinese Academy of Sciences.

\end{document}